\documentclass[twocolumn]{aastex62}
\usepackage{graphicx}
\usepackage{color}
\usepackage{amsmath}

\newcommand{\lrr}{Living Rev. Relativ.}
\newcommand{\prx}{Phys. Rev. X}
\newcommand{\aph}{Astropart. Phys.}

\begin{document}

\title{Search for Multi-messenger Sources of Gravitational Waves and High-energy Neutrinos \\ with Advanced LIGO during its first Observing Run, {\sc Antares} and IceCube}

\author{A.~Albert}
\affiliation{Universit\'e de Strasbourg, CNRS,  IPHC UMR 7178, F-67000 Strasbourg, France}
\author{M.~Andr\'e}
\affiliation{Technical University of Catalonia, Laboratory of Applied Bioacoustics, Rambla Exposici\'o, 08800 Vilanova i la Geltr\'u, Barcelona, Spain}
\author{M.~Anghinolfi}
\affiliation{INFN - Sezione di Genova, Via Dodecaneso 33, 16146 Genova, Italy}
\author{M.~Ardid}
\affiliation{Institut d'Investigaci\'o per a la Gesti\'o Integrada de les Zones Costaneres (IGIC) - Universitat Polit\`ecnica de Val\`encia. C/  Paranimf 1, 46730 Gandia, Spain}
\author{J.-J.~Aubert}
\affiliation{Aix Marseille Univ, CNRS/IN2P3, CPPM, Marseille, France}
\author{J.~Aublin}
\affiliation{APC, Univ Paris Diderot, CNRS/IN2P3, CEA/Irfu, Obs de Paris, Sorbonne Paris Cit\'e, France}
\author{T.~Avgitas}
\affiliation{APC, Univ Paris Diderot, CNRS/IN2P3, CEA/Irfu, Obs de Paris, Sorbonne Paris Cit\'e, France}
\author{B.~Baret}
\affiliation{APC, Univ Paris Diderot, CNRS/IN2P3, CEA/Irfu, Obs de Paris, Sorbonne Paris Cit\'e, France}
\author{J.~Barrios-Mart\'{\i}}
\affiliation{IFIC - Instituto de F\'isica Corpuscular (CSIC - Universitat de Val\`encia) c/ Catedr\'atico Jos\'e Beltr\'an, 2 E-46980 Paterna, Valencia, Spain}
\author{S.~Basa}
\affiliation{LAM - Laboratoire d'Astrophysique de Marseille, P\^ole de l'\'Etoile Site de Ch\^ateau-Gombert, rue Fr\'ed\'eric Joliot-Curie 38,  13388 Marseille Cedex 13, France}
% ...
\collaboration{(ANTARES collaboration)}
\thanks{Email: antares.spokesperson@in2p3.fr}
\author{M.~G.~Aartsen}
\affiliation{Dept.~of Physics and Astronomy, University of Canterbury, Private Bag 4800, Christchurch, New Zealand}
\author{M.~Ackermann}
\affiliation{DESY, D-15738 Zeuthen, Germany}
\author{J.~Adams}
\affiliation{Dept.~of Physics and Astronomy, University of Canterbury, Private Bag 4800, Christchurch, New Zealand}
\author{J.~A.~Aguilar}
\affiliation{Universit\'e Libre de Bruxelles, Science Faculty CP230, B-1050 Brussels, Belgium}
\author{M.~Ahlers}
\affiliation{Niels Bohr Institute, University of Copenhagen, DK-2100 Copenhagen, Denmark}
\author{M.~Ahrens}
\affiliation{Oskar Klein Centre and Dept.~of Physics, Stockholm University, SE-10691 Stockholm, Sweden}
\author{D.~Altmann}
\affiliation{Erlangen Centre for Astroparticle Physics, Friedrich-Alexander-Universit\"at Erlangen-N\"urnberg, D-91058 Erlangen, Germany}
\author{K.~Andeen}
\affiliation{Department of Physics, Marquette University, Milwaukee, WI, 53201, USA}
\author{T.~Anderson}
\affiliation{Dept.~of Physics, Pennsylvania State University, University Park, PA 16802, USA}
\author{I.~Ansseau}
\affiliation{Universit\'e Libre de Bruxelles, Science Faculty CP230, B-1050 Brussels, Belgium}
% ...
\collaboration{(IceCube Collaboration)}
\thanks{Email: analysis@icecube.wisc.edu}
\author{B.~P.~Abbott}
\affiliation{LIGO, California Institute of Technology, Pasadena, CA 91125, USA}
\author{R.~Abbott}
\affiliation{LIGO, California Institute of Technology, Pasadena, CA 91125, USA}
\author{T.~D.~Abbott}
\affiliation{Louisiana State University, Baton Rouge, LA 70803, USA}
\author{S.~Abraham}
\affiliation{Inter-University Centre for Astronomy and Astrophysics, Pune 411007, India}
\author{F.~Acernese}
\affiliation{Universit\`a di Salerno, Fisciano, I-84084 Salerno, Italy}
\affiliation{INFN, Sezione di Napoli, Complesso Universitario di Monte S.Angelo, I-80126 Napoli, Italy}
\author{K.~Ackley}
\affiliation{OzGrav, School of Physics \& Astronomy, Monash University, Clayton 3800, Victoria, Australia}
\author{C.~Adams}
\affiliation{LIGO Livingston Observatory, Livingston, LA 70754, USA}
\author{V.~B.~Adya}
\affiliation{Max Planck Institute for Gravitational Physics (Albert Einstein Institute), D-30167 Hannover, Germany}
\affiliation{Leibniz Universit\"at Hannover, D-30167 Hannover, Germany}
\author{C.~Affeldt}
\affiliation{Max Planck Institute for Gravitational Physics (Albert Einstein Institute), D-30167 Hannover, Germany}
\affiliation{Leibniz Universit\"at Hannover, D-30167 Hannover, Germany}
\author{M.~Agathos}
\affiliation{University of Cambridge, Cambridge CB2 1TN, United Kingdom}
% ...

\collaboration{(The LIGO Scientific Collaboration and the Virgo Collaboration)}
\thanks{Email: lsc-spokesperson@ligo.org}
\thanks{Email: virgo-spokesperson@ego-gw.it}

\begin{abstract}
Astrophysical sources of gravitational waves, such as binary neutron star and black hole mergers or core-collapse supernovae, can drive relativistic outflows, giving rise to non-thermal high-energy emission. High-energy neutrinos are signatures of such outflows. The detection of gravitational waves and high-energy neutrinos from common sources could help establish the connection between the dynamics of the progenitor and the properties of the outflow.  We searched for associated emission of gravitational waves and high-energy neutrinos from astrophysical transients with minimal assumptions using data from Advanced LIGO from its first observing run O1, and data from the {\sc Antares} and IceCube neutrino observatories from the same time period. We focused on candidate events whose astrophysical origin could not be determined from a single messenger. We found no significant coincident candidate, which we used to constrain the rate density of astrophysical sources dependent on their gravitational wave and neutrino emission processes.
\end{abstract}

\section{Introduction}

We have entered the era of regular gravitational-wave (GW) discoveries. Since 2015, Advanced LIGO \citep{2015CQGra..32g4001L} and Advanced Virgo \citep{TheVirgo:2014hva} have discovered GWs from multiple binary black hole mergers \citep{2016PhRvX...6d1015A,PhysRevLett.118.221101,2041-8205-851-2-L35,PhysRevLett.119.141101} and a binary neutron star (BNS) merger \citep{PhysRevLett.119.161101,2018arXiv180511579T} during Advanced LIGO's first two and Advanced Virgo's first observing periods. The rate of detections is expected to significantly increase in upcoming observation periods \citep{2016LRR....19....1A}.

High-energy neutrinos carry information about hadronic acceleration in astrophysical phenomena, such as accreting black holes and supernovae \citep{2002RPPh...65.1025H} and about the environment of the emission site (e.g., \citealt{2003PhRvD..68h3001R,2012PhRvD..86h3007B,2006JCAP...05..003L}). Several high-energy neutrino observatories carry out joint searches with GW and electromagnetic facilities. The primary facilities are the IceCube Neutrino Observatory (hereafter IceCube), a gigaton Cherenkov detector located in the ice at the South Pole \citep{Aartsen:2016_JINST}; the {\sc Antares} neutrino telescope (hereafter {\sc Antares}), a ten megaton scale underwater Cherenkov detector in the Mediterranean Sea \citep{2011NIMPA.656...11A}; and the Pierre Auger Cosmic Ray Observatory \citep{2015arXiv150201323T}. 

A quasi-diffuse high-energy neutrino flux of cosmic origin has been identified by the IceCube Neutrino Observatory \citep{2013PhRvL.111b1103A,2013Sci...342E...1I}, at a flux level consistent with the latest constraints by the {\sc Antares} neutrino detector \citep{2041-8205-853-1-L7}. Evidence of neutrino emission from the blazar TXS 0506+056 provides the strongest indication to date 
that at least a fraction of the cosmic neutrinos are produced in blazars \citep{eaat1378,IC147,2018arXiv180704309A}.

Neutrinos detected via charged current $\nu_\mu$ interactions can be reconstructed with an angular uncertainty $\lesssim 1^\circ$. Since the directions of GWs can be reconstructed to within tens to hundreds of square degrees, a joint GW+neutrino observation could significantly improve the localization of a GW source, making electromagnetic follow-up observations faster and more feasible. In addition, combining the GW and neutrino data allows us to identify candidates that would not otherwise be significant for either GW or neutrino data alone

No common sources of GWs and high-energy neutrinos have been identified so far. Until now, observational constraints for astrophysical source populations have only been derived using Initial LIGO and Virgo, and the partially completed IceCube and {\sc Antares} detectors \citep{2011PhRvL.107y1101B,2013JCAP...06..008A,2014PhRvD..90j2002A}. In addition, searches have been carried out for the neutrino counterparts of binary black hole mergers detected during Advanced LIGO's first \citep{2016PhRvD..93l2010A,2017PhRvD..96b2005A,2016ApJ...829L..34G,2016PhRvD..94l2007A,2016ApJ...830L..11A,2017ApJ...850...21A} and second observing runs \citep{2017arXiv171003020A,2017ApJ...850...21A}, and BNS merger GW170817/GRB\,170817A \citep{2041-8205-850-2-L35,2018ApJ...857L...4A}.

In this paper we present a multi-messenger search for common transient sources of GWs and high-energy neutrinos using GW data from Advanced LIGO's first observing run (O1) and neutrino data from both {\sc Antares} and IceCube. 

The paper is organized as follows. In Section \ref{sec:detectors}, we describe the GW and neutrino observatories, and the data used in this analysis. We also briefly introduce our multi-messenger search method. In Section \ref{sec:results}, we present the results of our combined search and the corresponding constraints on astrophysical populations. We present our conclusions in Section \ref{sec:conclusion}.

\section{Detectors and Data Analysis}
\label{sec:detectors}
\subsection{Advanced LIGO}

Advanced LIGO's O1 observing run started on September 12, 2015, and lasted until January 19, 2016. During this period, Advanced LIGO had an unprecedented sensitivity to GW transients, which led to the discovery of multiple astrophysical GW signals \citep{2016PhRvX...6d1015A}.

We used the data from Advanced LIGO's two detectors in Hanford, Washington and Livingston, Louisiana, to carry out a generic GW transient search, called {\it coherent WaveBurst} (cWB) \citep{2008CQGra..25k4029K,2011PhRvD..83j2001K,2016PhRvD..93d2004K}, using minimal assumptions on the source properties. We adopted the triggers from the all-sky, unmodeled, short duration, transient search reported by LIGO and Virgo~\cite{2016PhRvD..93l2004A}. In this, we quantified the significance of GW event candidates using a test statistic $\rho$ constructed in the framework of constrained maximum likelihood analysis \citep{2008CQGra..25k4029K}. We considered GW signal candidates with $\rho\geq6$, corresponding to a GW false alarm rate (FAR) $\mbox{FAR}_{\rm GW}\approx1$\,day$^{-1}$. Beyond $\rho$, cWB outputs the time of the GW candidate, as well as its directional probability distribution, or {\it skymap} \citep{2011PhRvD..83j2001K}. We calculate the GW skymap either up to its 90\% confidence region, or up to 320\,deg$^2$ divided into 2000 tiles of $0.4^\circ\times0.4^\circ$ size, whichever is smaller.

We assign each GW candidate one of three classifications, C1, C2, or C3, based on its time-frequency morphology \citep{2016PhRvD..93l2004A,2017PhRvD..95d2003A}.  These labels are assigned to help separate likely noise transients from other events. Candidates with frequency evolutions consistent with noise fluctuations often occurring in LIGO-Virgo data were placed into class C1. Multiple time-frequency morphologies were included. An example category are events for which at least $80\%$ GW energy is within a bandwidth of $5\,$Hz. Such a narrow band is characteristic of power and mechanical resonance lines in GW detectors. 

From the remaining candidates, those whose frequency increases with time, i.e. those similar in morphology to compact binary mergers, were placed in class C3. All other GW candidates were placed in class C2 \citep{2016PhRvD..93l2004A,2017PhRvD..95d2003A}. 

This grouping reduces the FAR for events within C2 and C3, without eliminating the chance of identifying a high-significance signal in C1. 

In this search we used the C2 and C3 classes together, which have a higher probability of being astrophysical, and treated the C1 class separately. We calculated the background distribution of the test statistic separately for these these two categories. For a given event, its GW p-value $p_{\rm GW}$ is calculated by comparing the reconstructed $\rho$ value to the background distribution of $\rho$ in the same category as the event. Because the C1 and C2+C3 searches are statistically independent, we include a trial factor of 2 in our final significance.

Overall, cWB identified 46 GW candidates during the $T_{\rm obs} = 48.6$\,days of coincident data from the LIGO Hanford and LIGO Livingston detectors, which is consistent with our background expectation. 23 of these candidates fell into the C1 category, while 23 were identified as C2+C3.

To characterize the background distribution of the ranking statistic $\rho$ for GW candidates, we carried out the same search over GW data after applying time shifts between the data from the two LIGO detectors, with time shifts much greater than travel time of GWs between the LIGO detectors (10\,ms). This technique ensures that no short GW transient appears simultaneously in the data streams of the two detectors, and is therefore able to characterize the performance of the search in the detector noise. We carried out the analysis over 500 different time shifts to collect a large background data set. We found a total of 23494 background GW candidates with $\rho\geq6$. A subset of 11005 of these were identified as C1, while 12489 were C2+C3. The FARs for C1 and C2+C3 are both $\sim0.5$\,day$^{-1}$.

\subsection{IceCube}

IceCube is a cubic-kilometer sized neutrino observatory~\citep{Aartsen:2016_JINST} installed in the ice at the geographic South Pole in Antarctica between depths of 1450\,m and 2450\,m. It is a gigaton-scale array of photosensors with a duty cycle higher than 99\%. IceCube observes neutrinos coming from all directions, but by using the Earth as a shield to block background cosmic ray-induced muons, it achieves very high detection efficiency for neutrinos originating in the Northern celestial hemisphere with energies above $\cal{O}$(1)\,TeV. Neutrinos originating in the Southern sky are detected with high efficiency above $\cal{O}$(100)\,TeV.

IceCube is sensitive to all neutrino flavors and both charged-current and neutral current interactions. For this search we focus on muon neutrinos that produce muons in charged-current interactions. These neutrinos are the most suitable for the search due to their superior angular reconstructions and high detection efficiency in the northern sky. 

We adopted a selection of through-going muons used in IceCube's online analyses \citep{1742-6596-718-6-062029,2016arXiv161206028I}, which follows an event selection similar to that used in point source searches \citep{2017ApJ...835..151A}. This event selection picks out primarily cosmic-ray-induced background events, with an expectation of 4.0 events in the northern sky (predominantly generated by atmospheric neutrinos) and 2.7 events in the southern sky (predominantly muons generated by high energy cosmic rays interactions in the atmosphere above the detector) per 1000 seconds.  

Between the beginning and the end of LIGO's O1 observing run, we identified 41985 neutrino candidates using IceCube's online analysis. The analysis determined the time of arrival, reconstructed energy, as well as the directional point spread function of each neutrino candidate.

\subsection{ANTARES}

The {\sc Antares}~neutrino telescope, located deep (2500\,m) in the Mediterranean Sea, 40\,km from Toulon, France, has been continuously operating since 2008. It is a 10\,megaton-scale array of photosensors, detecting neutrinos with energies above $\cal{O}$(100)\,GeV, with a duty cycle higher than $90\%$.

The selection criteria for the {\sc Antares} neutrino candidates were optimized based on the observed background rate and followed the same philosophy as the one used in the follow-up of GW170817~\citep{2041-8205-850-2-L35}. The events were selected from the most recent offline-reconstructed dataset, that incorporated dedicated calibrations, in terms of positioning~\citep{2012JInst...7T8002A}, timing~\citep{2011APh....34..539A} and efficiency~\citep{antareseff}. Only upgoing $\nu_{\mu}$ neutrino candidates, detected by their muon tracks, were considered in this analysis.

A time-dependent selection criterion, based on the quality of the muon track reconstruction, was optimized such that a selected high-energy neutrino event in a time window of 1000\,s and within the 90\% confidence contour of a GW would yield a significance of 3$\sigma$, i.e. have a probability of less than $2.7\times 10^{-3}$ of arising due to atmospheric backgrounds. We rely on a sample of simulated GW events~\citep{2014ApJ...795..105S} to extract a relationship between the signal-to-noise ratio of an event and the area of 90\% confidence region for the GW localization. This latter relation is used to extrapolate the size of the confidence region to sub-threshold GW events. This size is then convolved with the {\sc Antares} visible sky and its acceptance in local coordinates, to obtain the median 90\% confidence region of possible GW events.

In this specific study, the reduced time and space windows enable us to decrease the associated background, and therefore to relax the quality criteria that classify reconstructed tracks as upward going events. As a consequence the dominant background component is downgoing atmospheric muons misreconstructed as upgoing, hence mimicking neutrino-induced muons.

Each event is characterized by its detection time, arrival direction, directional uncertainty, and number of detected photons.  The latter is used here as an energy proxy.

The {\sc Antares} trigger rate varies with the environmental conditions, in particular the ambient background which is correlated with the sea current. Thus, using a time-dependent selection criterion instead of a constant value as used in point-source searches allows increasing the number of selected signal events. For an $E^{-2}$ spectrum the improvement is $45\% \pm 15\%$, depending on the time and data-taking conditions. This optimization improves the volume probed and correspondingly the number of detectable joint GW+high energy neutrino sources by the {\sc Antares} component of the joint analysis, by a factor 1.5 to 2.

With this new analysis, which considers the detector sensitivity at the time of the GW candidate, we obtain a total of 907 selected high-energy neutrino candidates with {\sc Antares} between the beginning and end of the O1 observation run, corresponding to an expected average of 0.1 neutrinos within a 1000\,s time window.

\subsection{Multi-messenger Analysis}

We jointly analyzed GW and neutrino event candidates to search for common sources using a multi-messenger search algorithm \citep{2012PhRvD..85j3004B}, which was already followed in a previous joint search \citep{2014PhRvD..90j2002A}. We used the significance of GW and neutrino candidates independently, as well as their temporal and directional coincidence, to quantify the significance of joint events. 

We adopted $\rho$ as the ranking statistic for GW candidates. We calculated the significance of GW candidate $i$ by calculating its $p$-value $p_{\rm GW,i}$ based on its $\rho_{\rm i}$ value, separately for the C1 and C2+C3 classes. That is, $p_{\rm GW,i}$ is defined as the fraction of background GW candidates with $\rho\geq\rho_{\rm i}$ and within the same signal category as GW candidate $i$.
For neutrino candidates, we used their reconstructed energy $\epsilon_{\rm \nu}$ as the ranking statistic. For {\sc Antares}, $\epsilon_{\rm \nu}$ is approximated with the number of detected photons corresponding to a given event, while for IceCube it is the energy reconstructed by the detection algorithm. We calculated the significance of neutrino candidate $j$ by calculating its $p$-value $p_{\rm \nu,j}$ based on the energy proxy $\epsilon_{\rm \nu,j}$. In the following for simplicity we will refer to this as the reconstructed energy. For IceCube, we considered all detected neutrino candidates within a declination band of $\pm5^\circ$ around the declination of candidate $j$. The candidate's $p$-value was then calculated as the fraction of background neutrino candidates within this band with energies $\epsilon_{\rm \nu}\geq\epsilon_{\rm \nu,j}$. This calculation accounts for the fact that the energy distribution for neutrino candidates in IceCube changes little with right ascension, but depends strongly on declination. For {\sc Antares}, $p_{\rm \nu,j}$ was calculated using Monte Carlo simulations as a probability of observing a neutrino energy $\epsilon_{\rm \nu}\geq\epsilon_{\rm \nu,j}$ given the observed neutrino direction.

In this analysis, temporal coincidence is a binary classification.  Any neutrino arriving within $\pm500$\,s of a GW candidate is considered temporally coincident \citep{2011APh....35....1B}.  Directional coincidence is quantified as the product of the GW skymap and neutrino reconstructed point spread function, marginalized over the whole sky.

In order to quantify the significance of joint event candidates, we carried out a Monte Carlo simulation to obtain their background distribution. One realization consisted of the following steps: (i) We randomly select a GW event candidate from the candidates identified in time-shifted GW data. (ii) We randomly select a neutrino candidate from the set of all observed neutrino candidates, and assign this to the selected GW candidate. We keep its original parameters, other than its time of arrival, which is changed to reflect the fact that we consider the two events to be temporally coincident. Importantly, we fix the neutrino's direction with respect to the neutrino detector's position, and calculate its right ascension and declination by assuming it arrived at the same time as the GW candidate it was assigned to. 

We realized 20000 times the steps described above both for the case of {\sc Antares} and for IceCube, and used these background simulations to calculate the $p$-value $p_{\rm sky}$ of directional coincidence. 

For neutrino candidates in temporal coincidence with GW candidates, we combined the three $p$-values from above into one ranking statistic $X^{2}$, following Fisher's method \citep{Fisher}:
\begin{equation}
X^{2} = -2\ln(p_{\rm GW}\cdot p_{\rm \nu}\cdot p_{\rm sky})
\end{equation}
For neutrino candidates not in such coincidence we assigned $X^{2}=0$. This results in a $X^{2}$ distribution with one component of positive values distributed according to the coincidence simulation described above, and one component located at zero. The fraction in the former component, i.e. the fraction of neutrino background events in GW coincidence, is $1-\mbox{Poiss}(0,\mbox{FAR}_{\rm GW}\Delta T)$. Here, Poiss$(k,\lambda)$ is the Poisson probability of observing $k$ events given $\lambda$ expected events, and $\Delta T = 1000$\,s is our search time window.

We quantified the significance of joint signal candidate $i$ using the $p$-value 
\begin{equation}
p^{(i)}_{\rm GW+\nu} = \int_{X^{2}_{\rm i}}^{\infty}p_{\rm BG}(X^{2'})dX^{2'},
\end{equation}
where $p_{\rm BG}(X^{2})$ is the distribution of $X^{2}$ for background events. Note that this $p$-value is defined for every neutrino candidate, also those not in temporal coincidence with a GW. For the latter category $p^{(i)}_{\rm GW+\nu}$.

A more detailed description of the method can be found in \cite{2012PhRvD..85j3004B}.

\subsection{Calculating population constraints}

The expected amplitude $h_{\rm rss}$ from a source depends on its distance $r$ as well as its total radiated GW energy $E_{\rm GW}$:
\begin{equation} 
h_{\rm rss}(E_{\rm GW},r) = \frac{\kappa G^{1/2}}{\pi c^{3/2}} \frac{E_{\rm GW}^{1/2}}{r f_{0}},
\end{equation}
where $c$ is the speed of light, $G$ is the gravitational constant, $f_0$ is the characteristic frequency of the GW, and $\kappa$ is an $O(1)$ dimensionless constant, which we take to be $(5/2)^{1/2}$ \citep{2013arXiv1304.0210S}. This value corresponds to a rotational GW source, such as a BNS merger or a rapidly rotating neutron star. 

We model the expected high-energy neutrino spectrum as $dn_{\nu}/dE_{\nu} = \Phi_0 E_{\nu}^{-2}$ within the energy band $E_{\rm \nu}\in[100\,\mbox{GeV},100\,\mbox{PeV}]$. For this model the neutrino spectral parameter $\Phi_0$ at Earth is  $\Phi_0 = E_{\rm \nu,iso}(4\pi r^2)^{-1}/6$, where $E_{\rm \nu,iso}$ is total isotropic-equivalent energy emitted in neutrinos. Combining $\Phi_0$ with the detectors' effective areas we can calculate the expected number of detected neutrinos $\langle N_{\nu}\rangle$. This in turn determines the probability that at least one neutrino will be detected from the source, given that it is beamed towards the observer:
\begin{equation}
p_{\rm det,\nu,X}(E_{\rm \nu,iso},r) = 1 - \mbox{Poiss}(0,\langle N_{\nu}\rangle_{\rm X}).
\end{equation}

Upon non-detection, we can obtain constraints on the population of GW+neutrino sources. Let $f_{\rm GW,IC}(h_{\rm rss})$ and $f_{\rm GW,A}(h_{\rm rss})$ be the fractions of GW+neutrino events with $h_{\rm rss}$ root-sum-squared GW strain amplitude that are expected to surpass a specific significance, here taken as that of our most significant event. Here and below, the subscript IC is used for IceCube and A for {\sc Antares}. We only consider the fraction of GW events here that have a temporally coincident neutrino candidate.

The rate upper limit $R_{\rm UL}$ of common sources will then be:
\begin{equation}
R_{\rm UL} = \frac{3.9f_{\rm b}}{T_{\rm obs}}\left[\int_{0}^{\infty} 4\pi r^2 p_{\rm det} dr \right]^{-1},
\label{eq:Rul}
\end{equation}
where $f_{\rm b}\equiv(1-\cos\theta_{\rm j})^{-1}$ is the neutrino emission's beaming factor for jet opening half angle $\theta_{\rm j}$, the factor 3.9 arises from the Poisson distribution and corresponds to a Neyman 90\% confidence-level upper limit, and 
\begin{equation}
\begin{split}
p_{\rm det} = f_{\rm GW,IC}\cdot p_{\rm det,\nu,IC} + f_{\rm GW,A}\cdot p_{\rm det,\nu,A} \\
- f_{\rm GW,IC}\cdot f_{\rm GW,A}\cdot p_{\rm det,\nu,IC}\cdot p_{\rm det,\nu,A}.
\label{eq:pdet}
\end{split}
\end{equation}
Here, the last term on the right side ensures that a simultaneous detection by IceCube and {\sc Antares} is not counted twice. 

\section{Results}
\label{sec:results}

We found that 42 of the 46 GW event candidates had temporally coincident neutrino candidates for IceCube, with a total of 195 coincident neutrinos. We identified no temporally coincident neutrino candidates for {\sc Antares}. These results are consistent with our background expectation.

None of the joint GW+neutrino candidates we identified have sufficiently high significance to consider them a detection. Our most significant event corresponds to a GW candidate recorded on December 18, 2015 at 11:40:17 UTC, and a neutrino candidate observed 296\,s later. There is a strong directional coincidence between the candidates, with $p_{\rm sky}=0.01$. The GW $p$-value for the event is $p_{\rm GW}=10^{-3}$. The GW candidate is classified as C2+C3. The neutrino candidate was detected at $(\mbox{R.A.},\mbox{Dec})=(312.5^\circ,-25.3^\circ)$. It had a reconstructed muon energy of $127.3$\,TeV. This is a typical energy for a background event in the southern sky, and corresponds to a neutrino $p$-value of $p_{\rm \nu}=0.43$. The $p$-value of our most significant event, considering the whole observing run, is $0.82$, making our results consistent with expectation from the background.

\begin{figure*}
\includegraphics[width=0.5\textwidth]{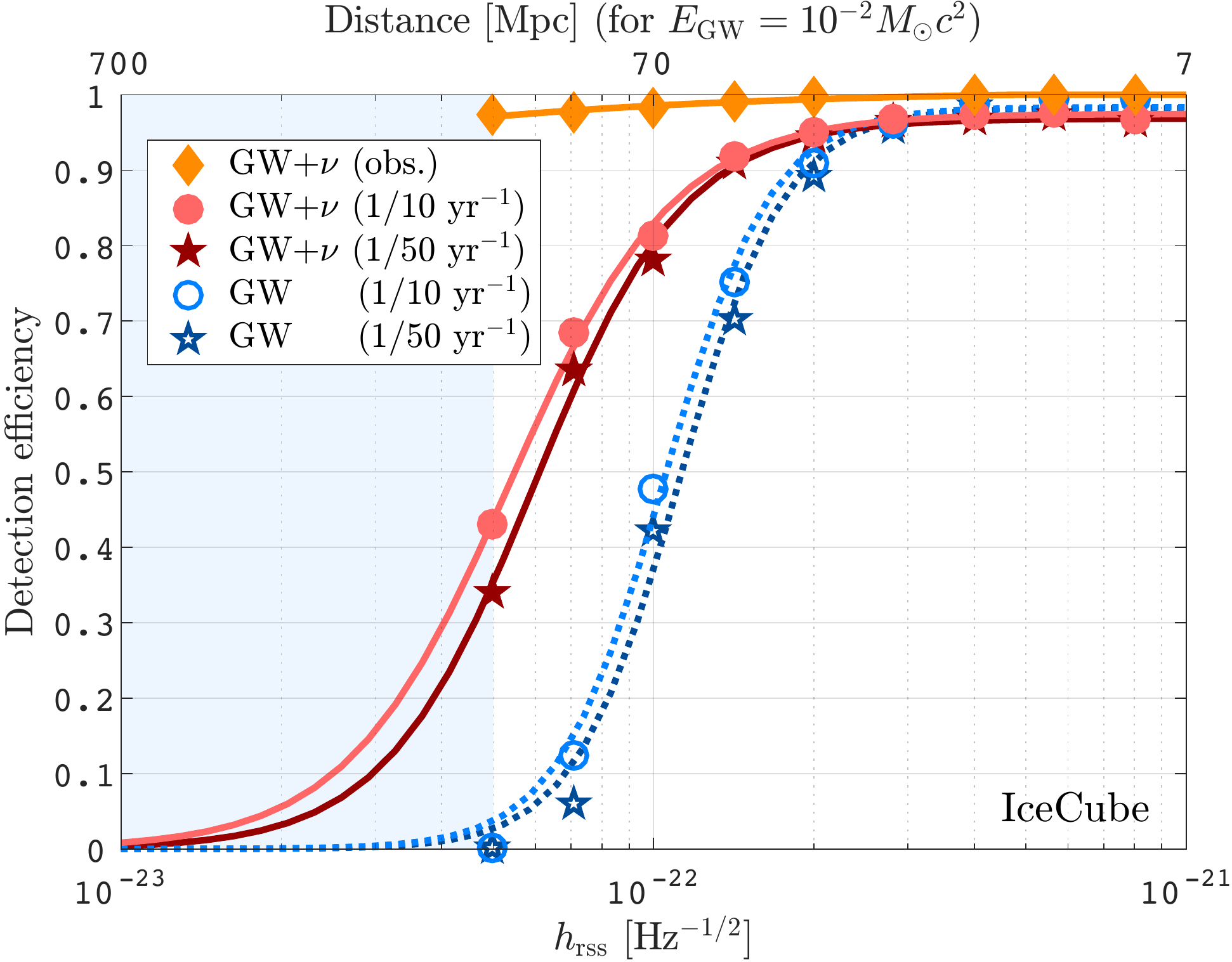}%
\includegraphics[width=0.5\textwidth]{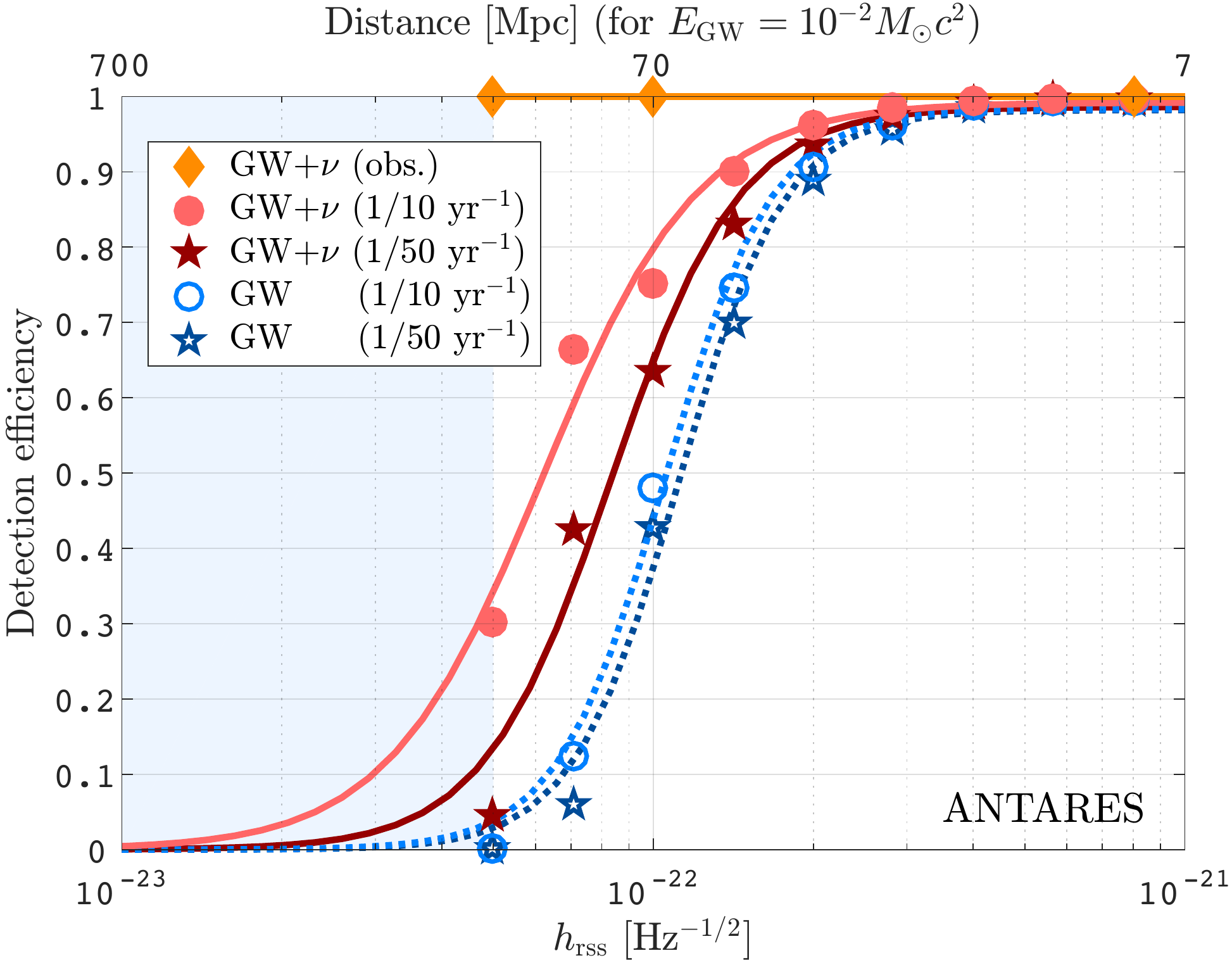}%
\caption{\label{fig:sensitivity_curve} Fraction of simulated astrophysical GW+neutrino events whose significance exceeds a threshold as a function of the GW $h_{\rm rss}$, assuming a sine-Gaussian gravitational waveform described in Section \ref{sec:sensitivity}.  Separate curves are shown for the cases of detections by IceCube+LIGO (left) and {\sc Antares}+LIGO (right). Results are shown for different significance thresholds, with thresholds set at the most significant event [GW$+\nu$\,(obs.)], as well as thresholds corresponding to FARs 1/10\,yr$^{-1}$ and 1/50\,yr$^{-1}$. For comparison, we further show results for GW-only searches, also for FARs 1/10\,yr$^{-1}$ and 1/50\,yr$^{-1}$. On the top of the figures we also show the source distance corresponding to $h_{\rm rss}$, assuming $E_{\rm GW}=10^{-2}$M$_\odot$c$^2$. Below $5\times10^{-23}$, we find that the GW search is unable to detect events (shaded area).}
\end{figure*}

\subsection{Sensitivity}
\label{sec:sensitivity}

We calculated the sensitivity of our search using simulated multi-messenger signals. We generated gravitational waveforms with varying amplitudes that we superimposed on the data. We adopted a sine-Gaussian gravitational waveform with characteristic frequency $f_0=153$\,Hz and quality factor $Q=9$. This standard waveform has been used for past searches, which allows comparison to prior results and the characterization of sensitivity (see, e.g., \citealt{2010PhRvD..81j2001A}). The sensitivity of GW detectors gradually decreases for frequencies away from the most sensitive band around $200$\,Hz. See \citep{2008CQGra..25d5002B} for a comparison of search sensitivities and \citep{2011PhRvD..83j2001K} for a comparison for localization accuracy for different gravitational waveforms. 

We used Monte Carlo simulations to generate a set of detected astrophysical high-energy neutrinos.  We draw the energies of the incoming neutrinos from a distribution of $dN_{\nu}/dE_{\nu}\propto E_{\nu}^{-2}$, consistent with the scaling expected for particle acceleration in relativistic jets \citep{1997PhRvL..78.2292W}. A softer spectrum, or the addition of a spectral cutoff, would make our resulting sensitivity somewhat weaker \citep{2016PhRvD..93l2010A}. We chose a lower limit for the neutrino energies of $300$\,GeV for IceCube and $100$\,GeV for {\sc Antares}.

We evaluated our search sensitivity as follows. For a given GW signal amplitude and assuming an astrophysical neutrino was detected from the source, we calculate the fraction of simulated GW+neutrino events that are reconstructed with $p_{\rm GW+\nu}$ below a threshold value. This gives us $f_{\rm GW,IC}(h_{\rm rss})$ and $f_{\rm GW,A}(h_{\rm rss})$, as defined earlier. We calculate these fractions for a range of GW signal amplitudes, characterized by the root-sum-squared GW strain $h_{\rm rss}$. We compute fractions for multiple thresholds: 

(i) First, we consider $p_{\rm GW+\nu}$ of our most significant event for IceCube. For {\sc Antares}, as there was no coincident GW+neutrino event, any coincidence by itself passes our threshold. 

(ii) We consider the expected most significant background events over 10\,yr and 50\,yr observation periods. To obtain these thresholds, we use Monte Carlo simulations to generate multiple realizations of 10\,yr and 50\,yr joint observation periods, and for each realization we find the event with the lowest $p_{\rm GW+\nu}$.

Fig. \ref{fig:sensitivity_curve} shows our search's detection efficiency as a function of $h_{\rm rss}$, separately for IceCube and {\sc Antares}, for different significance thresholds. We also show results for both GW+neutrino and GW-only sensitivities. For example, for $h_{\rm rss}=10^{-22}$\,Hz$^{-1/2}$ we find that 80\% of those GW+neutrino injections for which a neutrino is detected will have  FAR\,$<1/50$\,yr$^{-1}$, while only 43\% of GW events have FAR\,$<1/50$\,yr$^{-1}$. We also find that below $h_{\rm rss}=5\times 10^{-23}$\,Hz$^{-1/2}$ the GW search is unable to detect these events. 

We also see in Fig. \ref{fig:sensitivity_curve} how our sensitivity changes if instead of the most significant event of the present search we use as threshold a FAR of 1/10\,yr$^{-1}$ and 1/50\,yr$^{-1}$. For comparison, we also show the sensitivity curve for GW-only searches. We see that there is little difference between results for 1/10\,yr$^{-1}$ and 1/50\,yr$^{-1}$ FAR values, for either detector.

\subsection{Population constraints}

We used our non-detection to obtain constraints on the population of GW+neutrino sources. We carried out Monte Carlo simulations to compute the direction dependent effective area of the detectors, separately for IceCube and {\sc Antares}. Adopting a neutrino spectrum $E_{\nu}^{2} dn_{\nu}/dE_{\nu} = \Phi_0$, where $n_{\nu}$ is the neutrino fluence at the detector, we found that the sky-averaged expected number of detected neutrinos are $\langle N_{\nu}\rangle_{\rm IC} = 30(\Phi_0/\mbox{GeV\,cm}^{-2})$ and $\langle N_{\nu}\rangle_{\rm A} = 1.2(\Phi_0/\mbox{GeV\,cm}^{-2})$ for IceCube (IC) and {\sc Antares} (A), respectively. 

\begin{figure}
\includegraphics[width=0.5\textwidth]{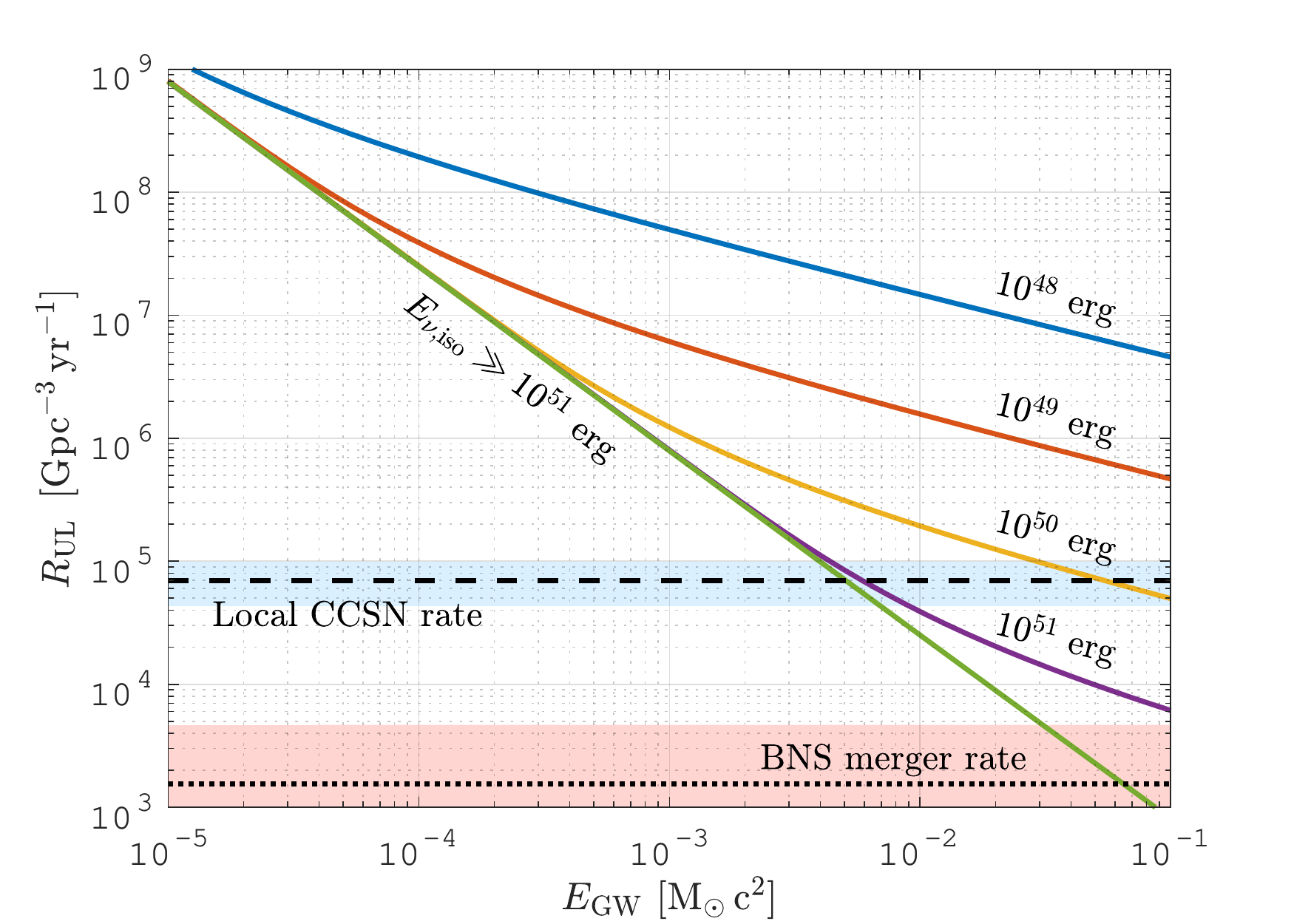}%
\caption{\label{fig:UL} Upper limits for the rate density of GW+neutrino sources as functions of $E_{\rm GW}$, for different values of $E_{\rm iso,\nu}$ (see numerical values of $E_{\rm iso,\nu}$ in the figure), for a sine-Gaussian gravitational waveform described in Section \ref{sec:sensitivity}. We assume a beaming factor $f_{\rm b} = 10$. For comparison, we show the rate density of local core-collapse supernovae (CCSNe; dashed line, rate error region shown in blue), and that of BNS mergers (dotted line, rate error region shown in red).}
\end{figure}

We used $f_{\rm GW,IC}(h_{\rm rss})$ and $f_{\rm GW,A}(h_{\rm rss})$ along with $\langle N_{\nu}\rangle$ to calculate $p_{\rm det}$ using Eq. \ref{eq:pdet}, which we substituted into Eq. \ref{eq:Rul} to obtain the population rate upper limit $R_{\rm UL}$. Fig. \ref{fig:UL} shows our results for $R_{\rm UL}$ for different source parameters. 

In Fig. \ref{fig:UL} we assume a beaming factor of $f_{\rm b}= 10$. The constraints linearly scale with $f_{\rm b}$. The expected beaming factor varies between sources. For low-luminosity gamma-ray bursts (GRBs), it can be as low as $f_{\rm b} \lesssim 14$ \citep{2007ApJ...662.1111L}. For long GRBs, typical jet opening angles are $\theta_{\rm j} = 3^\circ-10^\circ$, with some extending up to $\approx20^\circ$ \citep{2014ARA&A..52...43B}, corresponding to a beaming factor $f_{\rm b}=(1-\cos\theta_{\rm j})^{-1}=10-10^{3}$. 

Short GRBs were found to have comparable beaming factors based on their observed jet breaks and rate \citep{2014ARA&A..52...43B}. Nevertheless, the detection of GRB\,170817A at a higher observing angle of $\sim30^\circ\pm15^\circ$ \citep{2018arXiv180511579T} implied weaker effective beaming. Radio observations of the GRB's afterglow indicate that the outflow had a narrowly collimated relativistic jet with $\theta_{\rm j}<5^\circ$ as well as a broader, less energetic component \citep{2018arXiv180609693M,2018arXiv180800469G}. The origin of this structured outflow remains the subject of active debate \citep{2018arXiv180502870A,2017arXiv171203237L,2018Natur.554..207M,2018PTEP.2018d3E02I,2017arXiv171005896G,2017ApJ...848L..25H,2018arXiv180502870A,2018arXiv180207328V}. 

It is instructive to compare the present limits to previous results. Here we look at the latest estimates that used Initial LIGO-Virgo and the partially completed IceCube detector \citep{2014PhRvD..90j2002A}. Considering a fiducial source emission of $E_{\rm GW}=10^{-2}$\,M$_{\odot}$c$^{2}$ and $E_{\rm \nu,iso}=10^{51}$\,erg, assuming a beaming factor of $f_{\rm b}=10$, this previous search obtained a joint source rate upper limit of $1.1\times10^{7}$\,Gpc$^{-3}$yr$^{-1}$. The present search updates this constraint to $4\times10^{4}$\,Gpc$^{-3}$yr$^{-1}$, an improvement of more than 2 orders of magnitude.

\subsection{Discussion}

Here we briefly review the expected emission parameters of sources of interest, and compare the our rate density constraints to expectations. While our constraints take into account the total emitted energy in both GWs and high-energy neutrinos, and the high-energy beaming factor, the source constraints are also affected by the chosen gravitational waveform and the neutrino spectrum, which we do not explore here in detail. The comparison below should therefore be considered qualitative.

We show in Fig. \ref{fig:UL} the local ($z=0$) rate density of core-collapse supernovae (CCSNe) and BNS mergers. The rate of neutron star--black hole mergers, which also could produce relativistic jets, is expected to be lower, $\lesssim 50$\,Gpc$^{-3}$yr$^{-1}$ \citep{2017ApJ...849L..14G}. For CCSNe, it is possible that a large fraction of them drive relativistic jets \citep{2017arXiv170408298P}, potentially resulting in high-energy neutrino emission. Many of these jets may be stalled, however, before they are able to break through the stellar envelope \citep{2001PhRvL..87q1102M,2016PhRvD..93h3003S}. The resulting choked jets will have no observable gamma-ray emission, making high-energy neutrinos an interesting way to probe them. 

For CCSNe, we adopted the local rate of $7\pm 3 \times 10^{4}$\,Gpc$^{-3}$yr$^{-1}$ from \cite{2011MNRAS.412.1473L}. For BNS mergers, we adopted the rate $1540^{+3200}_{-1220}$\,Gpc$^{-3}$yr$^{-1}$ obtained from the detection of GW170817 \citep{PhysRevLett.119.161101}. 

The total energy emitted in GWs in BNS mergers is a few percent of a Solar mass. It depends on the neutron star masses as well as the nuclear equation of state \citep{2016PhRvD..94b4023B}. The expected rate of neutron star--black hole mergers falls below the shown range, while their GW energy could extend beyond $10^{-1}$\,M$_\odot$c$^2$, even for black hole masses $\lesssim 10$\,M$_\odot$ which can disrupt a neutron star upon merger. 

The range of $E_{\rm GW}$ is uncertain for CCSNe. Numerical simulations of stellar core collapse typically predict low GW emission, with $E_{\rm GW} \lesssim 10^{-7}$\,M$_\odot$c$^2$ \citep{2009CQGra..26f3001O,2013ApJ...766...43M,2010CQGra..27s4005Y,2012PTEP.2012aA301K}. For core-collapse events with rapidly rotating cores $E_{\rm GW}$ may be boosted to $10^{-2}$\,M$_\odot$c$^2$ if a substantial fraction of the newly formed protoneutron star rotational energy is radiated away in GWs \citep{2002ApJ...565..430F,2009ApJ...702.1171C,2013CQGra..30l3001B,2016ApJ...818...94K}. Fallback accretion onto the protoneutron star can further increase the available angular momentum for GW emission \citep{2012ApJ...761...63P}. 

High-energy neutrino emission from relativistic jets driven by either CCSNe or BNS mergers is not well understood. For GRBs, the total radiated energy $E_{\rm \nu,iso}$ can be comparable to the energy radiated in gamma-rays \citep{1997PhRvL..78.2292W}, although $E_{\rm \nu,iso}$ from GRBs has been observationally constrained by the non-detection of coincident neutrinos \citep{2012Natur.484..351I,2017ApJ...843..112A,2013A&A...559A...9A}. 

Neutrino emission can be enhanced for sub-photospheric dissipation processes, in which the observable gamma-ray flux is reduced by absorption \citep{2013PhRvL.110x1101B}. A particularly interesting scenario is emission while the jet is still inside the stellar envelope \citep{2001PhRvL..87q1102M,2003PhRvD..68h3001R,2012PhRvD..86h3007B,2016PhRvD..93h3003S,2016PhRvD..93e3010T}. As these events are faint or dark in gamma-rays, their $E_{\rm \nu,iso}$ is not strongly bound by observations as is the case for GRBs.

Recently, there has been significant interest in high-energy neutrino emission from BNS mergers. \cite{2017ApJ...848L...4K} found that the most promising neutrino sources are GRBs with extended emission which could produce $E_{\rm \nu,iso}\sim 10^{51}$\,erg. Extended emission refers to the weaker X-ray/gamma-ray emission observed for some short GRBs that follow the main short burst, which typically lasts for a hundred seconds. The origin of this emission is currently not understood. \cite{2017ApJ...849..153F} investigated the possibility that a long-lived neutron star remnant survives the BNS merger, and calculated the interaction between winds from the remnant with matter ejected from the merger. They found that this interaction could produce neutrinos over a period of weeks to a year that could reach $\sim10^{50}$\,erg energy. This particular emission model is not constrained by the present search due to its expected duration.

Following the discovery of BNS merger GW170817, \cite{2018MNRAS.476.1191B} looked at the expected neutrino flux for GRBs with structured jets observed at large viewing angles, finding a low $E_{\rm \nu,iso}\sim 10^{44}$\,erg. \cite{2018arXiv180511613K} studied neutrino emission in jets burrowing through the mildly relativistic ejecta of BNS mergers. They found that this trans-ejecta neutrino emission, when viewed on-axis, can reach $E_{\rm \nu,iso}\sim 10^{51}$\,erg.

Binary black hole mergers could also produce electromagnetic and neutrino emission if the black holes reside in a gaseous environment, although this scenario is not expected to arise for the majority of events. The first observational hint for such  was the observation of a possible short GRB by the Gamma-ray Burst Monitor on the Fermi satellite \citep{2016ApJ...826L...6C}. Scenarios that can result in electromagnetic and neutrino emission include mergers in the accretion disks of active galactic nuclei \citep{2017ApJ...835..165B,2017MNRAS.464..946S,2017NatCo...8..831B}, gas or debris remaining around the black holes from their prior evolution (\citealt{2016ApJ...821L..18P,2016ApJ...823L..29K,2016PhRvD..93l3011M,2017ApJ...839L...7D,2016ApJ...822L...9M}; but see \citealt{2017MNRAS.465.4406K}), and binary black hole formation inside a collapsing star (\citealt{2016ApJ...819L..21L}; but see \citealt{2017MNRAS.470L..92D}). The electromagnetic and neutrino brightness of binary black hole mergers within these scenarios is currently not well constrained. Continued follow-up observations of mergers discovered through GWs in the future will be able to confirm or provide interesting constraints on these models.

\section{Conclusion}
\label{sec:conclusion}

We searched for joint sources of GWs and high-energy neutrinos using observations from Advanced LIGO during its first observing run O1, and the {\sc Antares} and IceCube neutrino observatories. We identified no significant coincident GW and neutrino candidates.

We used the non-detection to obtain constraints on the rate density of multi-messenger GW+neutrino sources as functions of the energy emitted in gravitational waves and neutrinos. For realistic multi-messenger source rate densities of $<10^{5}$\,Gpc$^{-3}$yr$^{-1}$, the derived limits are constraining in the strong-emission regime of $E_{\rm GW}\gtrsim10^{-2}$\,M$_{\odot}$c$^2$ and $E_{\rm iso,\nu}\gtrsim10^{51}$\,erg. Such GW brightness is highly optimistic for CCSN events but it is more realistic for the case of compact binary mergers, while such neutrino brightness is comparable to the gamma-ray brightness of GRBs.

The considered observing period had an effective duration of just $\sim0.13$\,yr, which will be surpassed by future GW observing runs. In addition, we anticipate that LIGO's sensitivity will improve by a factor of $\sim 2$ upon reaching design sensitivity \citep{2016LRR....19....1A}.  Furthermore, other detectors such as Virgo will be operational in future observing periods (Virgo was partially operational during the second observing run, O2).  Meanwhile, planned next-generation neutrino detectors at the South Pole \citep{2014arXiv1412.5106I}, the Mediterranean \citep{2016JPhG...43h4001A} and in Lake Baikal \citep{2018arXiv180810353B} will lead to similarly significant improvements in sensitivity to high-energy astrophysical neutrinos. In light of these gains, we expect our sensitivity to possible multi-messenger GW+neutrino sources to improve significantly in the near future.

\begin{acknowledgments}
 The {\sc Antares} Collaboration acknowledge the financial support of the funding agencies:
Centre National de la Recherche Scientifique (CNRS), Commissariat \`a
l'\'ener\-gie atomique et aux \'energies alternatives (CEA),
Commission Europ\'eenne (FEDER fund and Marie Curie Program),
Institut Universitaire de France (IUF), IdEx program and UnivEarthS
Labex program at Sorbonne Paris Cit\'e (ANR-10-LABX-0023 and
ANR-11-IDEX-0005-02), Labex OCEVU (ANR-11-LABX-0060) and the
A*MIDEX project (ANR-11-IDEX-0001-02),
R\'egion \^Ile-de-France (DIM-ACAV), R\'egion
Alsace (contrat CPER), R\'egion Provence-Alpes-C\^ote d'Azur,
D\'e\-par\-tement du Var and Ville de La
Seyne-sur-Mer, France;
Bundesministerium f\"ur Bildung und Forschung
(BMBF), Germany; 
Istituto Nazionale di Fisica Nucleare (INFN), Italy;
Nederlandse organisatie voor Wetenschappelijk Onderzoek (NWO), the Netherlands;
Council of the President of the Russian Federation for young
scientists and leading scientific schools supporting grants, Russia;
National Authority for Scientific Research (ANCS), Romania;
Mi\-nis\-te\-rio de Econom\'{\i}a y Competitividad (MINECO):
Plan Estatal de Investigaci\'{o}n (refs. FPA2015-65150-C3-1-P, -2-P and -3-P, (MINECO/FEDER)), Severo Ochoa Centre of Excellence and MultiDark Consolider (MINECO), and Prometeo and Grisol\'{i}a programs (Generalitat
Valenciana), Spain; 
Ministry of Higher Education, Scientific Research and Professional Training, Morocco.
We also acknowledge the technical support of Ifremer, AIM and Foselev Marine
for the sea operation and the CC-IN2P3 for the computing facilities.
The IceCube Collaboration gratefully acknowledges the following support:
USA -- U.S. National Science Foundation-Office of Polar Programs,
U.S. National Science Foundation-Physics Division,
Wisconsin Alumni Research Foundation,
Center for High Throughput Computing (CHTC) at the University of Wisconsin-Madison,
Open Science Grid (OSG),
Extreme Science and Engineering Discovery Environment (XSEDE),
U.S. Department of Energy-National Energy Research Scientific Computing Center,
Particle astrophysics research computing center at the University of Maryland,
Institute for Cyber-Enabled Research at Michigan State University,
and Astroparticle physics computational facility at Marquette University;
Belgium -- Funds for Scientific Research (FRS-FNRS and FWO),
FWO Odysseus and Big Science programmes,
and Belgian Federal Science Policy Office (Belspo);
Germany -- Bundesministerium f\"ur Bildung und Forschung (BMBF),
Deutsche Forschungsgemeinschaft (DFG),
Helmholtz Alliance for Astroparticle Physics (HAP),
Initiative and Networking Fund of the Helmholtz Association,
Deutsches Elektronen Synchrotron (DESY),
and High Performance Computing cluster of the RWTH Aachen;
Sweden -- Swedish Research Council,
Swedish Polar Research Secretariat,
Swedish National Infrastructure for Computing (SNIC),
and Knut and Alice Wallenberg Foundation;
Australia -- Australian Research Council;
Canada -- Natural Sciences and Engineering Research Council of Canada,
Calcul Qu\'ebec, Compute Ontario, Canada Foundation for Innovation, WestGrid, and Compute Canada;
Denmark -- Villum Fonden, Danish National Research Foundation (DNRF);
New Zealand -- Marsden Fund;
Japan -- Japan Society for Promotion of Science (JSPS)
and Institute for Global Prominent Research (IGPR) of Chiba University;
Korea -- National Research Foundation of Korea (NRF);
Switzerland -- Swiss National Science Foundation (SNSF).
The LIGO Scientific Collaboration and the Virgo Collaboration gratefully acknowledge the support of the United States
National Science Foundation (NSF) for the construction and operation of the
LIGO Laboratory and Advanced LIGO as well as the Science and Technology Facilities Council (STFC) of the
United Kingdom, the Max-Planck-Society (MPS), and the State of
Niedersachsen/Germany for support of the construction of Advanced LIGO
and construction and operation of the GEO600 detector.
Additional support for Advanced LIGO was provided by the Australian Research Council.
The authors gratefully acknowledge the Italian Istituto Nazionale di Fisica Nucleare (INFN),
the French Centre National de la Recherche Scientifique (CNRS) and
the Foundation for Fundamental Research on Matter supported by the Netherlands Organisation for Scientific Research,
for the construction and operation of the Virgo detector
and the creation and support  of the EGO consortium.
The authors also gratefully acknowledge research support from these agencies as well as by
the Council of Scientific and Industrial Research of India,
the Department of Science and Technology, India,
the Science \& Engineering Research Board (SERB), India,
the Ministry of Human Resource Development, India,
the Spanish  Agencia Estatal de Investigaci\'on,
the Vicepresid\`encia i Conselleria d'Innovaci\'o, Recerca i Turisme and the Conselleria d'Educaci\'o i Universitat del Govern de les Illes Balears,
the Conselleria d'Educaci\'o, Investigaci\'o, Cultura i Esport de la Generalitat Valenciana,
the National Science Centre of Poland,
the Swiss National Science Foundation (SNSF),
the Russian Foundation for Basic Research,
the Russian Science Foundation,
the European Commission,
the European Regional Development Funds (ERDF),
the Royal Society,
the Scottish Funding Council,
the Scottish Universities Physics Alliance,
the Hungarian Scientific Research Fund (OTKA),
the Lyon Institute of Origins (LIO),
the Paris \^{I}le-de-France Region,
the National Research, Development and Innovation Office Hungary (NKFI),
the National Research Foundation of Korea,
Industry Canada and the Province of Ontario through the Ministry of Economic Development and Innovation,
the Natural Science and Engineering Research Council Canada,
the Canadian Institute for Advanced Research,
the Brazilian Ministry of Science, Technology, Innovations, and Communications,
the International Center for Theoretical Physics South American Institute for Fundamental Research (ICTP-SAIFR),
the Research Grants Council of Hong Kong,
the National Natural Science Foundation of China (NSFC),
the Leverhulme Trust,
the Research Corporation,
the Ministry of Science and Technology (MOST), Taiwan
and
the Kavli Foundation.
The authors gratefully acknowledge the support of the NSF, STFC, MPS, INFN, CNRS and the
State of Niedersachsen/Germany for provision of computational resources.

\end{acknowledgments}


\begin{thebibliography}{}
\expandafter\ifx\csname natexlab\endcsname\relax\def\natexlab#1{#1}\fi
\providecommand{\url}[1]{\href{#1}{#1}}

\bibitem[{{Aab} {et~al.}(2015)}]{2015arXiv150201323T}
{Aab}, A., {et~al.} 2015, Nucl. Instrum. Methods Phys. Res A, 798, 172

\bibitem[{{Aab} {et~al.}(2016)}]{2016PhRvD..94l2007A}
---. 2016, Phys. Rev. D, 94, 122007

\bibitem[{{Aartsen} {et~al.}(2013{\natexlab{a}})}]{2013PhRvL.111b1103A}
{Aartsen}, M., {et~al.} 2013{\natexlab{a}}, Phys. Rev. Lett., 111, 021103

\bibitem[{{Aartsen} {et~al.}(2013{\natexlab{b}})}]{2013Sci...342E...1I}
---. 2013{\natexlab{b}}, Science, 342, 1242856

\bibitem[{{Aartsen} {et~al.}(2014{\natexlab{a}})}]{2014PhRvD..90j2002A}
---. 2014{\natexlab{a}}, Phys. Rev. D, 90, 102002

\bibitem[{{Aartsen} {et~al.}(2014{\natexlab{b}})}]{2014arXiv1412.5106I}
---. 2014{\natexlab{b}}, arXiv:1412.5106

\bibitem[{{Aartsen} {et~al.}(2017{\natexlab{a}})}]{Aartsen:2016_JINST}
---. 2017{\natexlab{a}}, JINST, 12, P03012

\bibitem[{{Aartsen} {et~al.}(2017{\natexlab{b}})}]{2016arXiv161206028I}
---. 2017{\natexlab{b}}, Astropar. Phys., 92, 30

\bibitem[{{Aartsen} {et~al.}(2017{\natexlab{c}})}]{2017ApJ...835..151A}
---. 2017{\natexlab{c}}, \apj, 835, 151

\bibitem[{{Aartsen} {et~al.}(2017{\natexlab{d}})}]{2017ApJ...843..112A}
---. 2017{\natexlab{d}}, \apj, 843, 112

\bibitem[{{Aartsen} {et~al.}(2018{\natexlab{a}})}]{eaat1378}
---. 2018{\natexlab{a}}, Science, 361.
\newblock \url{http://science.sciencemag.org/content/361/6398/eaat1378}

\bibitem[{{Aartsen} {et~al.}(2018{\natexlab{b}})}]{IC147}
---. 2018{\natexlab{b}}, Science, 361, 147.
\newblock \url{http://science.sciencemag.org/content/361/6398/147}

\bibitem[{{Abadie} {et~al.}(2010)}]{2010PhRvD..81j2001A}
{Abadie}, J., {et~al.} 2010, Phys. Rev. D, 81, 102001

\bibitem[{{Abadie} {et~al.}(2015)}]{2015CQGra..32g4001L}
---. 2015, Class. Quantum Grav., 32, 074001

\bibitem[{{Abbasi} {et~al.}(2012)}]{2012Natur.484..351I}
{Abbasi}, R., {et~al.} 2012, \nat, 484, 351

\bibitem[{{Abbott} {et~al.}(2016{\natexlab{a}})}]{2016PhRvX...6d1015A}
{Abbott}, B., {et~al.} 2016{\natexlab{a}}, \prx, 6, 041015

\bibitem[{{Abbott} {et~al.}(2016{\natexlab{b}})}]{2016PhRvD..93l2004A}
---. 2016{\natexlab{b}}, Phys. Rev. D, 93, 122004

\bibitem[{{Abbott} {et~al.}(2017{\natexlab{a}})}]{PhysRevLett.118.221101}
---. 2017{\natexlab{a}}, Phys. Rev. Lett., 118, 221101.
\newblock \url{https://link.aps.org/doi/10.1103/PhysRevLett.118.221101}

\bibitem[{{Abbott} {et~al.}(2017{\natexlab{b}})}]{2041-8205-851-2-L35}
---. 2017{\natexlab{b}}, \apjl, 851, L35.
\newblock \url{http://stacks.iop.org/2041-8205/851/i=2/a=L35}

\bibitem[{{Abbott} {et~al.}(2017{\natexlab{c}})}]{PhysRevLett.119.141101}
---. 2017{\natexlab{c}}, Phys. Rev. Lett., 119, 141101.
\newblock \url{https://link.aps.org/doi/10.1103/PhysRevLett.119.141101}

\bibitem[{{Abbott} {et~al.}(2017{\natexlab{d}})}]{PhysRevLett.119.161101}
---. 2017{\natexlab{d}}, Phys. Rev. Lett., 119, 161101.
\newblock \url{https://link.aps.org/doi/10.1103/PhysRevLett.119.161101}

\bibitem[{{Abbott} {et~al.}(2017{\natexlab{e}})}]{2017PhRvD..95d2003A}
---. 2017{\natexlab{e}}, Phys. Rev. D, 95, 042003

\bibitem[{{Abbott} {et~al.}(2018{\natexlab{a}})}]{2018arXiv180511579T}
---. 2018{\natexlab{a}}, arXiv:1805.11579

\bibitem[{{Abbott} {et~al.}(2018{\natexlab{b}})}]{2016LRR....19....1A}
---. 2018{\natexlab{b}}, \lrr, 21, 3.
\newblock \url{https://doi.org/10.1007/s41114-018-0012-9}

\bibitem[{{Abe} {et~al.}(2016)}]{2016ApJ...830L..11A}
{Abe}, K., {et~al.} 2016, \apjl, 830, L11

\bibitem[{{Abe} {et~al.}(2018)}]{2018ApJ...857L...4A}
---. 2018, \apjl, 857, L4

\bibitem[{{Acernese} {et~al.}(2015)}]{TheVirgo:2014hva}
{Acernese}, F., {et~al.} 2015, Class. Quantum Grav., 32, 024001

\bibitem[{{Adri{\'a}n-Mart{\'{\i}}nez} {et~al.}(2012)}]{2012JInst...7T8002A}
{Adri{\'a}n-Mart{\'{\i}}nez}, S., {et~al.} 2012, J. Instrum., 7, T08002

\bibitem[{{Adri{\'a}n-Mart{\'{\i}}nez}
  {et~al.}(2013{\natexlab{a}})}]{2013JCAP...06..008A}
---. 2013{\natexlab{a}}, \jcap, 6, 008

\bibitem[{{Adri{\'a}n-Mart{\'{\i}}nez}
  {et~al.}(2013{\natexlab{b}})}]{2013A&A...559A...9A}
---. 2013{\natexlab{b}}, \aap, 559, A9

\bibitem[{{Adri{\'a}n-Mart{\'{\i}}nez}
  {et~al.}(2016{\natexlab{a}})}]{2016PhRvD..93l2010A}
---. 2016{\natexlab{a}}, Phys. Rev. D, 93, 122010

\bibitem[{{Adri{\'a}n-Mart{\'{\i}}nez}
  {et~al.}(2016{\natexlab{b}})}]{2016JPhG...43h4001A}
---. 2016{\natexlab{b}}, J. Phys. G, 43, 084001

\bibitem[{{Ageron} {et~al.}(2011)}]{2011NIMPA.656...11A}
{Ageron}, M., {et~al.} 2011, Nucl. Instr. Meth. Phys. Res. A, 656, 11

\bibitem[{{Agostini} {et~al.}(2017)}]{2017ApJ...850...21A}
{Agostini}, M., {et~al.} 2017, \apj, 850, 21

\bibitem[{{Aguilar} {et~al.}(2007)}]{antareseff}
{Aguilar}, J., {et~al.} 2007, Nucl. Instrum. Meth. A, 570, 107

\bibitem[{{Aguilar} {et~al.}(2011)}]{2011APh....34..539A}
---. 2011, \aph, 34, 539

\bibitem[{{Albert} {et~al.}(2017{\natexlab{a}})}]{2017PhRvD..96b2005A}
{Albert}, A., {et~al.} 2017{\natexlab{a}}, Phys. Rev. D, 96, 022005

\bibitem[{{Albert} {et~al.}(2017{\natexlab{b}})}]{2017arXiv171003020A}
---. 2017{\natexlab{b}}, European Physical Journal C, 77, 911

\bibitem[{{Albert} {et~al.}(2017{\natexlab{c}})}]{2041-8205-850-2-L35}
---. 2017{\natexlab{c}}, \apjl, 850, L35.
\newblock \url{http://stacks.iop.org/2041-8205/850/i=2/a=L35}

\bibitem[{{Albert} {et~al.}(2018{\natexlab{a}})}]{2041-8205-853-1-L7}
---. 2018{\natexlab{a}}, \apjl, 853, L7.
\newblock \url{http://stacks.iop.org/2041-8205/853/i=1/a=L7}

\bibitem[{{Albert} {et~al.}(2018{\natexlab{b}})}]{2018arXiv180704309A}
---. 2018{\natexlab{b}}, arXiv:1807.04309 (to appear in ApJ. Lett)

\bibitem[{Alexander {et~al.}(2018)}]{2018arXiv180502870A}
Alexander, K.~D., {et~al.} 2018, \apjl, 863, L18.
\newblock \url{http://stacks.iop.org/2041-8205/863/i=2/a=L18}

\bibitem[{{Avrorin} {et~al.}(2018){Avrorin}, {Avrorin}, {Aynutdinov},
  {Bannash}, {Belolaptikov}, {Brudanin}, {Budnev}, {Doroshenko}, {Domogatsky},
  {Dvornick{\'y}}, {Dyachok}, {Dzhilkibaev}, {Fajt}, {Fialkovsky}, {Gafarov},
  {Golubkov}, {Gres}, {Honz}, {Kebkal}, {Kebkal}, {Khramov}, {Kolbin},
  {Konischev}, {Korobchenko}, {Koshechkin}, {Kozhin}, {Kulepov}, {Kuleshov},
  {Milenin}, {Mirgazov}, {Osipova}, {Panfilov}, {Pan'kov}, {Petukhov},
  {Pliskovsky}, {Rozanov}, {Rjabov}, {Rushay}, {Safronov}, {Simkovic},
  {Shoibonov}, {Solovjev}, {Sorokovikov}, {Shelepov}, {Suvorova}, {Shtekl},
  {Tabolenko}, {Tarashansky}, {Yakovlev}, {Zagorodnikov}, \&
  {Zurbanov}}]{2018arXiv180810353B}
{Avrorin}, A.~D., {Avrorin}, A.~V., {Aynutdinov}, V.~M., {et~al.} 2018,
  arXiv:1808.10353

\bibitem[{{Baret} {et~al.}(2011)}]{2011APh....35....1B}
{Baret}, B., {et~al.} 2011, \aph, 35, 1

\bibitem[{{Baret} {et~al.}(2012)}]{2012PhRvD..85j3004B}
---. 2012, Phys. Rev. D, 85, 103004

\bibitem[{{Bartos} {et~al.}(2013{\natexlab{a}}){Bartos}, {Beloborodov},
  {Hurley}, \& {M{\'a}rka}}]{2013PhRvL.110x1101B}
{Bartos}, I., {Beloborodov}, A.~M., {Hurley}, K., \& {M{\'a}rka}, S.
  2013{\natexlab{a}}, Phys. Rev. Lett., 110, 241101

\bibitem[{{Bartos} {et~al.}(2013{\natexlab{b}}){Bartos}, {Brady}, \&
  {M{\'a}rka}}]{2013CQGra..30l3001B}
{Bartos}, I., {Brady}, P., \& {M{\'a}rka}, S. 2013{\natexlab{b}}, Class.
  Quantum Grav., 30, 123001

\bibitem[{{Bartos} {et~al.}(2012){Bartos}, {Dasgupta}, \&
  {M{\'a}rka}}]{2012PhRvD..86h3007B}
{Bartos}, I., {Dasgupta}, B., \& {M{\'a}rka}, S. 2012, Phys. Rev. D, 86, 083007

\bibitem[{{Bartos} {et~al.}(2011){Bartos}, {Finley}, {Corsi}, \&
  {M{\'a}rka}}]{2011PhRvL.107y1101B}
{Bartos}, I., {Finley}, C., {Corsi}, A., \& {M{\'a}rka}, S. 2011, Phys. Rev.
  Lett., 107, 251101

\bibitem[{{Bartos} {et~al.}(2017{\natexlab{a}}){Bartos}, {Kocsis}, {Haiman}, \&
  {M{\'a}rka}}]{2017ApJ...835..165B}
{Bartos}, I., {Kocsis}, B., {Haiman}, Z., \& {M{\'a}rka}, S.
  2017{\natexlab{a}}, \apj, 835, 165

\bibitem[{{Bartos} {et~al.}(2017{\natexlab{b}})}]{2017NatCo...8..831B}
{Bartos}, I., {et~al.} 2017{\natexlab{b}}, Nature Communications, 8, 831

\bibitem[{{Beauville} {et~al.}(2008)}]{2008CQGra..25d5002B}
{Beauville}, F., {et~al.} 2008, Class. Quantum Grav., 25, 045002

\bibitem[{{Berger}(2014)}]{2014ARA&A..52...43B}
{Berger}, E. 2014, \araa, 52, 43

\bibitem[{{Bernuzzi} {et~al.}(2016)}]{2016PhRvD..94b4023B}
{Bernuzzi}, S., {et~al.} 2016, Phys. Rev. D, 94, 024023

\bibitem[{{Biehl} {et~al.}(2018){Biehl}, {Heinze}, \&
  {Winter}}]{2018MNRAS.476.1191B}
{Biehl}, D., {Heinze}, J., \& {Winter}, W. 2018, \mnras, 476, 1191

\bibitem[{{Connaughton} {et~al.}(2016)}]{2016ApJ...826L...6C}
{Connaughton}, V., {et~al.} 2016, \apjl, 826, L6

\bibitem[{{Corsi} \& {M{\'e}sz{\'a}ros}(2009)}]{2009ApJ...702.1171C}
{Corsi}, A., \& {M{\'e}sz{\'a}ros}, P. 2009, \apj, 702, 1171

\bibitem[{{Dai} {et~al.}(2017){Dai}, {McKinney}, \&
  {Miller}}]{2017MNRAS.470L..92D}
{Dai}, L., {McKinney}, J.~C., \& {Miller}, M.~C. 2017, \mnras, 470, L92

\bibitem[{{de Mink} \& {King}(2017)}]{2017ApJ...839L...7D}
{de Mink}, S.~E., \& {King}, A. 2017, \apjl, 839, L7

\bibitem[{{Fang} \& {Metzger}(2017)}]{2017ApJ...849..153F}
{Fang}, K., \& {Metzger}, B.~D. 2017, \apj, 849, 153

\bibitem[{Fisher(1925)}]{Fisher}
Fisher, R. 1925, Statistical Methods for Research Workers (Oliver and Boyd,
  Edinburgh)

\bibitem[{{Fryer} {et~al.}(2002){Fryer}, {Holz}, \&
  {Hughes}}]{2002ApJ...565..430F}
{Fryer}, C.~L., {Holz}, D.~E., \& {Hughes}, S.~A. 2002, \apj, 565, 430

\bibitem[{{Gando} {et~al.}(2016)}]{2016ApJ...829L..34G}
{Gando}, A., {et~al.} 2016, \apjl, 829, L34

\bibitem[{{Ghirlanda} {et~al.}(2018)}]{2018arXiv180800469G}
{Ghirlanda}, G., {et~al.} 2018, arXiv:1808.00469

\bibitem[{{Gottlieb} {et~al.}(2018){Gottlieb}, {Nakar}, {Piran}, \&
  {Hotokezaka}}]{2017arXiv171005896G}
{Gottlieb}, O., {Nakar}, E., {Piran}, T., \& {Hotokezaka}, K. 2018, \mnras,
  479, 588

\bibitem[{{Gupta} {et~al.}(2017){Gupta}, {Arun}, \&
  {Sathyaprakash}}]{2017ApJ...849L..14G}
{Gupta}, A., {Arun}, K.~G., \& {Sathyaprakash}, B.~S. 2017, \apjl, 849, L14

\bibitem[{{Haggard} {et~al.}(2017)}]{2017ApJ...848L..25H}
{Haggard}, D., {et~al.} 2017, \apjl, 848, L25

\bibitem[{{Halzen} \& {Hooper}(2002)}]{2002RPPh...65.1025H}
{Halzen}, F., \& {Hooper}, D. 2002, Rep. Prog. Phys, 65, 1025

\bibitem[{{Ioka} \& {Nakamura}(2018)}]{2018PTEP.2018d3E02I}
{Ioka}, K., \& {Nakamura}, T. 2018, Progr. Theor. Exp. Phys., 2018, 043E02

\bibitem[{{Kashiyama} {et~al.}(2016){Kashiyama}, {Murase}, {Bartos}, {Kiuchi},
  \& {Margutti}}]{2016ApJ...818...94K}
{Kashiyama}, K., {Murase}, K., {Bartos}, I., {Kiuchi}, K., \& {Margutti}, R.
  2016, \apj, 818, 94

\bibitem[{{Kimura} {et~al.}(2018){Kimura}, {Murase}, {Bartos}, {Ioka}, {Heng},
  \& {M{\'e}sz{\'a}ros}}]{2018arXiv180511613K}
{Kimura}, S.~S., {Murase}, K., {Bartos}, I., {et~al.} 2018, arXiv:1805.11613

\bibitem[{{Kimura} {et~al.}(2017{\natexlab{a}}){Kimura}, {Murase},
  {M{\'e}sz{\'a}ros}, \& {Kiuchi}}]{2017ApJ...848L...4K}
{Kimura}, S.~S., {Murase}, K., {M{\'e}sz{\'a}ros}, P., \& {Kiuchi}, K.
  2017{\natexlab{a}}, \apjl, 848, L4

\bibitem[{{Kimura} {et~al.}(2017{\natexlab{b}}){Kimura}, {Takahashi}, \&
  {Toma}}]{2017MNRAS.465.4406K}
{Kimura}, S.~S., {Takahashi}, S.~Z., \& {Toma}, K. 2017{\natexlab{b}}, \mnras,
  465, 4406

\bibitem[{Kintscher {et~al.}(2016)}]{1742-6596-718-6-062029}
Kintscher, T., {et~al.} 2016, JPCS, 718, 062029.
\newblock \url{http://stacks.iop.org/1742-6596/718/i=6/a=062029}

\bibitem[{{Klimenko} {et~al.}(2008){Klimenko}, {Yakushin}, {Mercer}, \&
  {Mitselmakher}}]{2008CQGra..25k4029K}
{Klimenko}, S., {Yakushin}, I., {Mercer}, A., \& {Mitselmakher}, G. 2008,
  Class. Quantum Grav., 25, 114029

\bibitem[{{Klimenko} {et~al.}(2011){Klimenko}, {Vedovato}, {Drago}, {Mazzolo},
  {Mitselmakher}, {Pankow}, {Prodi}, {Re}, {Salemi}, \&
  {Yakushin}}]{2011PhRvD..83j2001K}
{Klimenko}, S., {Vedovato}, G., {Drago}, M., {et~al.} 2011, Phys. Rev. D, 83,
  102001

\bibitem[{{Klimenko} {et~al.}(2016)}]{2016PhRvD..93d2004K}
{Klimenko}, S., {et~al.} 2016, Phys. Rev. D, 93, 042004

\bibitem[{{Kotake} {et~al.}(2012){Kotake}, {Sumiyoshi}, {Yamada}, {Takiwaki},
  {Kuroda}, {Suwa}, \& {Nagakura}}]{2012PTEP.2012aA301K}
{Kotake}, K., {Sumiyoshi}, K., {Yamada}, S., {et~al.} 2012, Progr. Theor. Exp.
  Phys, 2012, 01A301

\bibitem[{{Kotera} \& {Silk}(2016)}]{2016ApJ...823L..29K}
{Kotera}, K., \& {Silk}, J. 2016, \apjl, 823, L29

\bibitem[{{Lazzati} {et~al.}(2018){Lazzati}, {Perna}, {Morsony},
  {Lopez-Camara}, {Cantiello}, {Ciolfi}, {Giacomazzo}, \&
  {Workman}}]{2017arXiv171203237L}
{Lazzati}, D., {Perna}, R., {Morsony}, B.~J., {et~al.} 2018, Phys. Rev. Lett.,
  120, 241103

\bibitem[{{Li} {et~al.}(2011){Li}, {Chornock}, {Leaman}, {Filippenko},
  {Poznanski}, {Wang}, {Ganeshalingam}, \& {Mannucci}}]{2011MNRAS.412.1473L}
{Li}, W., {Chornock}, R., {Leaman}, J., {et~al.} 2011, \mnras, 412, 1473

\bibitem[{{Liang} {et~al.}(2007){Liang}, {Zhang}, {Virgili}, \&
  {Dai}}]{2007ApJ...662.1111L}
{Liang}, E., {Zhang}, B., {Virgili}, F., \& {Dai}, Z.~G. 2007, \apj, 662, 1111

\bibitem[{{Loeb}(2016)}]{2016ApJ...819L..21L}
{Loeb}, A. 2016, \apjl, 819, L21

\bibitem[{{Loeb} \& {Waxman}(2006)}]{2006JCAP...05..003L}
{Loeb}, A., \& {Waxman}, E. 2006, JCAP, 5, 003

\bibitem[{{M{\'e}sz{\'a}ros} \& {Waxman}(2001)}]{2001PhRvL..87q1102M}
{M{\'e}sz{\'a}ros}, P., \& {Waxman}, E. 2001, Physical Review Letters, 87,
  171102

\bibitem[{{Moharana} {et~al.}(2016)}]{2016PhRvD..93l3011M}
{Moharana}, R., {et~al.} 2016, \prd, 93, 123011

\bibitem[{{Mooley} {et~al.}(2018{\natexlab{a}})}]{2018arXiv180609693M}
{Mooley}, K., {et~al.} 2018{\natexlab{a}}, Nature, 561, 355.
\newblock \url{https://doi.org/10.1038/s41586-018-0486-3}

\bibitem[{{Mooley} {et~al.}(2018{\natexlab{b}})}]{2018Natur.554..207M}
{Mooley}, K.~P., {et~al.} 2018{\natexlab{b}}, Nature, 554, 207

\bibitem[{{M{\"u}ller} {et~al.}(2013){M{\"u}ller}, {Janka}, \&
  {Marek}}]{2013ApJ...766...43M}
{M{\"u}ller}, B., {Janka}, H.-T., \& {Marek}, A. 2013, \apj, 766, 43

\bibitem[{{Murase} {et~al.}(2016){Murase}, {Kashiyama}, {M{\'e}sz{\'a}ros},
  {Shoemaker}, \& {Senno}}]{2016ApJ...822L...9M}
{Murase}, K., {Kashiyama}, K., {M{\'e}sz{\'a}ros}, P., {Shoemaker}, I., \&
  {Senno}, N. 2016, \apjl, 822, L9

\bibitem[{{Ott}(2009)}]{2009CQGra..26f3001O}
{Ott}, C.~D. 2009, Class. Quantum Grav., 26, 063001

\bibitem[{{Perna} {et~al.}(2016){Perna}, {Lazzati}, \&
  {Giacomazzo}}]{2016ApJ...821L..18P}
{Perna}, R., {Lazzati}, D., \& {Giacomazzo}, B. 2016, \apjl, 821, L18

\bibitem[{{Piran} {et~al.}(2017){Piran}, {Nakar}, {Mazzali}, \&
  {Pian}}]{2017arXiv170408298P}
{Piran}, T., {Nakar}, E., {Mazzali}, P., \& {Pian}, E. 2017, arXiv:1704.08298

\bibitem[{{Piro} \& {Thrane}(2012)}]{2012ApJ...761...63P}
{Piro}, A.~L., \& {Thrane}, E. 2012, \apj, 761, 63

\bibitem[{{Razzaque} {et~al.}(2003){Razzaque}, {M{\'e}sz{\'a}ros}, \&
  {Waxman}}]{2003PhRvD..68h3001R}
{Razzaque}, S., {M{\'e}sz{\'a}ros}, P., \& {Waxman}, E. 2003, Phys. Rev. D, 68,
  083001

\bibitem[{{Senno} {et~al.}(2016){Senno}, {Murase}, \&
  {M{\'e}sz{\'a}ros}}]{2016PhRvD..93h3003S}
{Senno}, N., {Murase}, K., \& {M{\'e}sz{\'a}ros}, P. 2016, Phys. Rev. D, 93,
  083003

\bibitem[{{Singer} {et~al.}(2014)}]{2014ApJ...795..105S}
{Singer}, L., {et~al.} 2014, \apj, 795, 105

\bibitem[{{Stone} {et~al.}(2017){Stone}, {Metzger}, \&
  {Haiman}}]{2017MNRAS.464..946S}
{Stone}, N.~C., {Metzger}, B.~D., \& {Haiman}, Z. 2017, \mnras, 464, 946

\bibitem[{{Sutton}(2013)}]{2013arXiv1304.0210S}
{Sutton}, P.~J. 2013, arXiv:1304.0210

\bibitem[{{Tamborra} \& {Ando}(2016)}]{2016PhRvD..93e3010T}
{Tamborra}, I., \& {Ando}, S. 2016, Phys. Rev. D, 93, 053010

\bibitem[{{Veres} {et~al.}(2018)}]{2018arXiv180207328V}
{Veres}, P., {et~al.} 2018, arXiv:1802.07328

\bibitem[{{Waxman} \& {Bahcall}(1997)}]{1997PhRvL..78.2292W}
{Waxman}, E., \& {Bahcall}, J. 1997, Phys. Rev. Lett, 78, 2292

\bibitem[{{Yakunin} {et~al.}(2010){Yakunin}, {Marronetti}, {Mezzacappa},
  {Bruenn}, {Lee}, {Chertkow}, {Hix}, {Blondin}, {Lentz}, {Messer}, \&
  {Yoshida}}]{2010CQGra..27s4005Y}
{Yakunin}, K.~N., {Marronetti}, P., {Mezzacappa}, A., {et~al.} 2010, Class.
  Quantum Grav., 27, 194005

\end{thebibliography}
\end{document}